\documentclass{acmart}
\AtBeginDocument{%
  \providecommand\BibTeX{{%
    \normalfont B\kern-0.5em{\scshape i\kern-0.25em b}\kern-0.8em\TeX}}}

\copyrightyear{2022}
\acmYear{2022}
\setcopyright{acmlicensed}
\acmConference[FAccT '22]{2022 ACM Conference on Fairness, Accountability, and Transparency}{June 21--24, 2022}{Seoul, Republic of Korea}
\acmBooktitle{2022 ACM Conference on Fairness, Accountability, and Transparency (FAccT '22), June 21--24, 2022, Seoul, Republic of Korea}
\acmPrice{15.00}
\acmDOI{10.1145/3531146.3533226}
\acmISBN{978-1-4503-9352-2/22/06}

\begin{document}

\title[Demographic-Reliant Algorithmic Fairness]{Demographic-Reliant Algorithmic Fairness: Characterizing the Risks of Demographic Data Collection in the Pursuit of Fairness}

\author{McKane Andrus}
\authornote{Both authors contributed equally to this research.}
\email{mckane@partnershiponai.org}
\author{Sarah Villeneuve}
\authornotemark[1]
\email{sarah.v@partnershiponai.org}
\affiliation{%
  \institution{Partnership on AI}
  \city{San Francisco}
  \state{California}
  \country{USA}
}

% \author{Anonymo}

\renewcommand{\shortauthors}{Andrus and Villeneuve}

\begin{abstract}

 Most proposed algorithmic fairness techniques require access to demographic data in order to make performance comparisons and standardizations across groups, however this data is largely unavailable in practice, hindering the widespread adoption of algorithmic fairness. Through this paper, we consider calls to collect more data on demographics to enable algorithmic fairness and challenge the notion that discrimination can be overcome with smart enough technical methods and sufficient data. We show how these techniques largely ignore broader questions of data governance and systemic oppression when categorizing individuals for the purpose of fairer algorithmic processing. In this work, we explore under what conditions demographic data should be collected and used to enable algorithmic fairness methods by characterizing a range of social risks to individuals and communities. For the risks to individuals we consider the unique privacy risks of sensitive attributes, the possible harms of miscategorization and misrepresentation, and the use of sensitive data beyond data subjects' expectations. Looking more broadly, the risks to entire groups and communities include the expansion of surveillance infrastructure in the name of fairness, misrepresenting and mischaracterizing what it means to be part of a demographic group, and ceding the ability to define what constitutes biased or unfair treatment. We argue that, by confronting these questions before and during the collection of demographic data, algorithmic fairness methods are more likely to actually mitigate harmful treatment disparities without reinforcing systems of oppression. Towards this end, we assess privacy-focused methods of data collection and use and participatory data governance structures as proposals for more responsibly collecting demographic data.

\end{abstract}





\maketitle

\section{Introduction} \label{sec:intro}
% Algorithmic decision-making has been widely accepted as a novel approach to overcoming the purported cognitive and subjective limitations of human decision makers by providing “objective” data-driven recommendations. Yet, a

As organizations increasingly adopt algorithmic decision-making systems (ADMS) for their efficiency and purported objectivity, harmful and discriminatory decision patterns have been observed in contexts such as healthcare \cite{obermeyer_dissecting_2019}, hiring \cite{chen_investigating_2018,seyyed-kalantari_chexclusion_2021,raghavan_mitigating_2019}, criminal justice \cite{washington_how_2018}, and education \cite{rimfeld_-level_2020}. In response, a swath of algorithmic fairness strategies have been proposed to better understand how ADMS treat certain individuals and groups in order to detect and mitigate harmful biases. 

Most current algorithmic fairness techniques require access to demographic data (such as race, gender, or sexuality) in order to make performance comparisons and standardizations across groups \cite{veale_fairer_2017}. These demographic-based algorithmic fairness techniques look to overcome discrimination and social inequality with novel metrics operationalizing notions of fairness and by collecting the requisite data, often removing broader questions of governance and politics from the equation \cite{balayn_beyond_2021}. This paper problematizes this approach, arguing instead that collecting more data in support of fairness is not always the answer and can actually exacerbate or introduce harm for marginalized individuals and groups. We believe more discussion is needed in the machine learning community around the consequences of “fairer” algorithmic decision-making, and what conditions must be satisfied in order responsibly collect and use demographic data for fairness purposes.

The rest of the paper proceeds as follows: Sections \ref{sec:disag} and \ref{sec:measurement} outline the importance of demographic data for addressing algorithmic discrimination and the challenges practitioners currently face when attempting to collect and use this data for fairness purposes. In Sections \ref{sec:indivrisk} and \ref{sec:commrisks} we characterize the risks that can emerge through the collection and use of demographic data first to individuals and then to communities at large. Finally, in Section \ref{sec:paths} we outline promising approaches to mitigating a number of the risks discussed in this paper. 

% In doing so, we hope that readers will better understand the affordances and limitations of using demographic data to detect and mitigate discrimination in institutional decision-making more broadly.

\section{Defining Demographic Data}
Throughout the literature on algorithmic fairness, we see a range of terms used to refer to data about demographics, such as \textit{demographic attributes}, \textit{protected categories/classes/characteristics}, and \textit{sensitive features}, among others. In most cases, these terms are used to refer simply to the variables in a dataset for which different fairness metrics should be computed, without much nuance regarding what is actually captured by the variable. Concepts such as gender, race, ethnicity, and sexuality are collapsed into single categories that are seen as self-evident, inherent characteristics of one's identity. This approach is understandable, seeing as most of the points of reference regarding anti-discrimination, or fairness more broadly, in machine learning stem from legal notions of discrimination that prescribe standards of fairness using census-like data \cite{barocas_big_2016}. Single variable representations of gender, race, ethnicity, and other demographic categories are schemas of social categorization that are both highly measurable and highly actionable, as they tell a relatively clear story about what types of differences exists across groups. What they often miss, however, is how these schemas are themselves products of social systems and conceal parts of the story when taken at face value -- to be Black in the United States does not just mean having the "attribute" of dark skin, it also means being on the receiving end of generations of oppression and disenfranchisement through social, governmental, and political means.

If not just a demographic variable in a dataset and not just an individual characteristic, how then should we think about the concepts underlying demographic data like race or gender? In this work, we draw from long histories of scholarship that interrogate categorization schemas and the social harms that can stem from their uncritical adoption \cite{guyan2022queer,bowker_sorting_1999,dembroff_real_2018,hacking_looping_1995,hanna_towards_2020,hu_what_2020,keyes_counting_2019,zuberi_white_2008}. \textit{Stemming from these understandings, we see demographic data, and demographic categories more broadly, as an attempt to collapse complex social concepts into categorical variables based on observable or self-identifiable characteristics.} While data of this kind can certainly help establish claims of unfairness, as we explore in many of the sections below, the mismatch between efforts to make these categories legible to computers and the actual, multidimensonal, and often fluid nature of class-membership can undermine work around fairness from the start.

\section{Algorithmic Fairness, Disaggregated Evaluations, and Demographic Data} \label{sec:disag}
As algorithmic decision-making systems become more widespread, there is greater risk for the systems to reinforce historical inequalities and engender new forms of discrimination in ways that are difficult to assess. In many cases, when ADMS discriminate against protected groups, they do so indirectly \cite{zhang_causal_2017}. While it is certainly possible for machine learning systems to base decisions off of features like race, more often the tools uncover trends and correlations that effectively discriminate across groups without relying on demographic variables.

When looking at how algorithms can discriminate, it is important to consider the different ways in which bias can enter the picture. The most often discussed point of entry is the data used to build the system. Biases in the data collection process and existing social inequalities will dictate the types of correlation that can be utilized by a machine learning system. If a group is underrepresented in the dataset or if the dataset embeds the results of historical discrimination and oppression in the form of biased features, it is to be expected that ADMS will have worse performance for or undervalue certain groups \cite{ntoutsi_bias_2020, olteanu_social_2019}.
% \footnote{For a detailed discussion of the many kinds of data bias, see \citet{ntoutsi_bias_2020} and \citet{olteanu_social_2019}.}

How ADMS are designed and towards what kinds of objectives, on the other hand, can have a large bearing on how discriminatory their outcomes are \cite{passi_problem_2019}. If optimizing for a goal that is poorly defined, or even discriminatorily defined, it is likely that a system will reproduce historical inequity and discrimination, just under a guise of objectivity and disinterestedness \cite{selbst_fairness_2019}. For example, the UK higher education admission algorithm that attempted to define aptitude as a combination of a predicted performance and secondary school quality systematically biased the outcomes for those coming from poorer or less-established secondary schools \cite{rimfeld_-level_2020}. Similarly, ADMS that ignore contextual differences between groups in an attempt to treat everyone equally often lead to discriminatory outcomes, such as in the case of hate speech detection systems that do not consider the identities of the speaker \cite{davidson_automated_2017,davidson_racial_2019}.

Though the types of discrimination discussed here represent a small subset of the myriad ways that ADMS can discriminate, they still bring up a difficult question — how should practitioners assess the potential discriminatory impacts of their systems? The nascent field of Algorithmic Fairness has contributed a number of strategies for identifying and even mitigating discrimination by ADMS, however, almost all of the proposed methods require the use of datasets which include the potentially discriminated against demographic attributes. Even outside of Algorithmic Fairness, conducting any kind of quantitative evaluation that disaggregates results across groups is likely to require data on group membership \cite{barocas_designing_2021}.

As previous research has highlighted, data on demographic categories is often unavailable due to a range of organizational challenges, legal barriers, and practical concerns \cite{andrus_what_2021}. Some privacy laws, such as the EU’s General Data Protection Regulation (GDPR), effectively only allow the collection of sensitive data such as race, religion, and sexuality under strict conditions of meaningful consent from data subjects \cite{goodman2016step,uk_information_commissioners_office_what_2020}. Some corporate privacy policies and standards, such as Privacy By Design, call for organizations to be intentional with their data collection practices, only collecting data they require and can specify a use for \cite{andrus_what_2021}. Given the uncertainty around whether or not it is acceptable to ask users and customers for their sensitive demographic information, most legal and policy teams urge their corporations to err on the side of caution and not collect these types of data unless legally required to do so. As a result, concerns over privacy often take precedence over ensuring product fairness since the trade-offs between mitigating bias and ensuring individual or group privacy are unclear \cite{andrus_what_2021}. 

Furthermore, prior work has shown that demographic data is generally only collected once a narrow, enforceable definition of discrimination is codified into law or corporate standards \cite{bogen_awareness_2020}. As such, the issue of missing demographic data is often only confronted and explicitly addressed once assessment and/or enforcement efforts begin in earnest\footnote{See, for example, the new data collection efforts mandated by \href{https://www.federalregister.gov/documents/2021/01/25/2021-01753/advancing-racial-equity-and-support-for-underserved-communities-through-the-federal-government}{Exec. Order No. 13985} and \href{https://www.federalregister.gov/documents/2021/06/30/2021-14127/diversity-equity-inclusion-and-accessibility-in-the-federal-workforce}{Exec. Order No. 14035}}. Even then, however, we see that anti-discrimination standards and practices vary widely across domains, and in many cases specific types of discrimination are legally sanctioned (e.g., “actuarial fairness” in insurance quotes and “legitimate aims” in employment law \cite{ochigame_beyond_2018}). Most often, however, legal anti-discrimination frameworks consider ignoring or omitting demographic attributes altogether non-discriminatory \cite{xiang2020reconciling}. When ADMS use this approach, often called “fairness through unawareness” or (in cases involving race) “color-blindness,” the results have often been shown to be just as discriminatory as whatever came before algorithmic decision-making \cite{kusner_counterfactual_2017}.  In other cases still, antidiscrimination law and policies can indirectly inhibit corporations from using demographic data, even if it is permitted, since being made aware of discrimination opens the door to legal liability if the discrimination is uncovered without a plan to successfully mitigate it \cite{andrus_what_2021}.

Beyond uncovering bias and discrimination, access to demographic data can help provide justification for the adequate representation and participation of various groups during the design and implementation of ADMS. The trajectory of COVID-19 data collection in the U.S. serves as a good example of this -- though the CDC requested racial demographic data to be collected on everyone who was treated for symptoms of COVID-19, racial demographics were frequently omitted in most local and state data collection efforts \cite{banco_cdc_2021}. As such, the unique vulnerabilities of Black, Indigenous, and Latinx individuals and communities against the virus were largely obscured until data collection and inference methods improved \cite{banco_cdc_2021}.

Partially as a result of this general absence of demographic data, we frequently see a cycle of ADMS development and deployment, exposure of egregious discrimination through individual reports, and then ad hoc system redesigns. \footnote{The AI Incident Database \cite{partnership_on_ai_welcome_2022} includes many examples of this, such as Google just removing the 'gorilla' tag after it was applied to Black users' photos \cite{hern_googles_2018} and Amazon scrapping their resume screening tool after it was shown to penalize experience with organizations such as "Women Who Code" \cite{grossman_amazon_2018}.} Without access to demographic data, it is difficult to assess these types of shortcomings before system deployment, and even after deployment it is likely that more insidious forms of discrimination remain hidden.

In the face of such a cycle, organizations looking to assess algorithmic fairness techniques have called for guidance on how to responsibly collect and use demographic data. However, prescribing adherence to statistical definitions of fairness on algorithmic systems without accounting for the social, economic, and political systems in which they are embedded can fail to benefit marginalized groups and undermine fairness efforts \cite{mitchell_algorithmic_2020,bakalar_fairness_2021}. Therefore, developing guidance requires a deeper understanding of the risks and trade-offs inherent to the use of demographic data. Efforts to detect and mitigate harms must account for the wider contexts and power structures that algorithmic systems, and the data that they draw on, are embedded in. 

\section{Common Concerns around Measuring Demographics} \label{sec:measurement}

When considering the risks of demographic data collection and use, it is often important to consider how the data is collected. Generally speaking, some combination of self-identification, ascription, and inference are used to create datasets that include demographics or to supplement existing datasets with demographic categories. Self-identification is usually operationalized as self-classification, or having data subjects select relevant categories from a set of options, ascription, also referred to as labelling, relies on data labelers or other second parties to determine the data subjects' demographics from existing data sources such as images or text, and inference or imputation use statistical methods and machine learning to guess subjects' attributes based on correlations found in datasets that already included demographic variables. Each of these approaches comes with trade-offs around privacy, data quality, and technical and economic feasibility that must be balanced when deciding whether to, and if so how to, collect demographic data \cite{andrus_what_2021}.

To start, each of these approaches takes on a unique set of risks to personal privacy. Self-identification arguably incurs the least risk, as requiring data subject participation ensures that they have more control over and awareness of what types of data about them exists. That being said, it can also have the impact of making people more aware of their privacy and increase concerns about what the data will be used for, an outcome many corporations try to avoid. Ascription and inference, on the other hand, allow data holders to define aspects about data subjects without giving them a say or even without them being made aware. These methods can differ in risk in that some individuals might be more comfortable having attributes like their gender identity predicted by an algorithm instead of say a data labeler given the growing disillusionment and dismissal of algorithmic profiling \cite{buchi_making_2021}. Furthermore, while ascription and inference can perhaps mitigate contests over privacy on the front end, they greatly increase the risk of public backlash if or when it is revealed that this data was collected, such as in the case of Facebook inferring "ethnic affinity" or "multicultural affinity" \cite{keegan_facebook_2021}.

Each measurement technique also comes with unique challenges around the quality of the resulting datasets. While self-reporting likely results in the most accurate labels, it can also produce the sparsest datasets, as many individuals will not share sensitive attributes unless they are sufficiently incentivized or bought into the goal of collection \cite{king2019becoming,andrus_what_2021}. In cases where ascription is used, on the other hand, datasets are more complete but much less accurate. Depending on the types of data available when ascribing attributes, whether it be images, written text, or some other source of metadata about a person, certain demographic categories may just not be at all determinable (e.g. sexuality when the available data is just images of faces). Inference techniques encounter a similar tradeoff of completeness for accuracy, but they are more scalable and so find more use in practice. Inference is also commonly used as a supplemental technique, allowing practitioners to fill in demographic attribute variables in incomplete datasets. While this is a low-cost strategy that can enable many kinds of algorithmic fairness analysis, it requires practitioners to be mindful of how inferred attributes introduce new sources of bias to the analysis \cite{chen_fairness_2019}.

Though concerns around privacy and data quality are deeply pervasive, costs and organizational risks are the most likely barriers to collecting demographic data \cite{andrus_what_2021}. In the case of self-identification, for many companies asking users or clients for their demographics would be seen as questionable if not outright nefarious. As such, the effort needed to communicate how demographic data would be used as well as the potential legal risks from privacy and anti-discrimination law simply set too high of a cost on collecting demographics through self-identification. Ascription, on the other hand, is generally cost-prohibitive for large datasets because it relies on paid employees and contractors to produce demographic information. That being said, it can be used to create small, high quality datasets for assessing system bias in cases where discrimination is likely to be based on ascribed characteristics, such as perceived gender or race \cite{basu_measuring_2020}. Inferential methods, despite carrying the most risk of inaccuracy when used by themselves, are extremely common in practice largely because of their immediate feasibility. In many domains it is only explicitly asking individuals for their sensitive attributes that is legally prohibited, not the actual use of those attributes to assess discrimination. As such, practitioners looking to assess their systems for discriminatory behavior are pushed to create what are known as \textit{proxy models} that predict sensitive attributes to assess potential discrimination \cite{kelley_anti-discrimination_2021,chen_fair_2018,bureau_using_2014}. Without explicit guidance or requirements around algorithmic antidiscrimination, this type of measurement strategy is likely to become more pervasive despite incurring some of the greatest risks to accuracy and privacy.

% Organizations also face difficulty capturing unobserved characteristics, such as disability, sexuality, and religion, as these categories are frequently missing and often unmeasurable \cite{tomasev_fairness_2021}. 

\section{Individual Risks of of Demographic Data Collection and Use} \label{sec:indivrisk}
When discussing the social risks of collecting demographic data, most researchers and practitioners focus mainly on the threats to individual privacy. In this section we expand on the individual privacy conversation and consider two more sources of risk -- individual misrepresentation and use beyond intended consent. Our goal with this section and the next is not to suggest that demographic data should never be used, but rather to build out a clearer picture of what future data collection efforts should attempt to address in their pursuit to enable less discriminatory decision making.

\subsection{Encroachments on Individual Privacy} \label{sec:indiv_privacy}

Though privacy is a commonly held concern when it comes to any type of data collection, the collection of demographic data requires special care and consideration. Sharing or otherwise determining sensitive attributes can expose individuals to various forms of direct or indirect harm, especially already marginalized and vulnerable individuals. Though there are numerous proposed methods for ensuring the privacy and security of sensitive attributes, the strategies for assessing (let alone mitigating) fairness or discrimination under privacy constraints are still very experimental and not commonly used \cite{farrand_neither_2020,jagielski_differentially_2019,kuppam_fair_2020}. As such, most efforts to collect sensitive demographic data will at some point in the pipeline require tying the data to individuals, necessarily risking individuals’ privacy.

One clear privacy risk of obtaining an individual’s demographic data is that this data can still be the basis for many types of discrimination. Though many countries have laws against direct discrimination, it is still a common occurrence due to the difficulty of proving discrimination in individual cases, especially algorithmic ones \cite{wachter_bias_2021}. In domains such as hiring \cite{quillian_meta-analysis_2017,quillian_evidence_2020}, advertising \cite{cabanas_does_2021,datta_automated_2015}, and pricing \cite{hupperich_empirical_2018,mikians_crowd-assisted_2013}, direct forms of discrimination, algorithmically mediated or not, are relatively common. For sectors like advertising, discriminatory practices are often justified by claims that differential treatment results in more helpful services, which may in fact be true. However, in a recent survey study of Facebook users, most were still uncomfortable with sensitive attributes being used as the basis for decisions around what they are being shown \cite{cabanas_does_2021}. 

In the most pernicious cases, demographic data can be used as the criteria for various forms of state or societally enacted violence, such as detainment and deportation based on documentation status in the United States. Even in cases where the targeted attribute (e.g., documentation status) is not collected, other accessible forms of data (e.g., country of birth and spoken language, contact lists) can be used to help infer the targeted attribute \cite{tomasev_fairness_2021}. As corporately collected data becomes increasingly requested by and made available to state agencies \cite{leetaru_facebook_2018,rozenshtein_surveillance_2018,tau_federal_2020}, it is critical that practitioners consider what types of identity-based violence individuals might be exposed to by providing self-categorizations.

Depending on what types of threat individuals' feel behind having their sensitive attributes revealed, it is also possible that the collection of demographic data can have a "chilling effect" on members of the groups most at-risk of discrimination or targeting. Once cognizant of the possibility that a platform or system is directly asking for or inferring demographic attributes, individuals may change their behavior on a platform or with a system to prevent being labeled or out of concerns such as "stereotype threat," the concern of being viewed as an example of a negative stereotype about some aspect of one's identity \cite{walton_expandable_2012}. A commonly suggested approach to reducing these forms of direct targeting risk is to “anonymize” or “de-identify” datasets, but even with these strategies marginalized individuals can still be vulnerable to "re-identification" \cite{chang_privacy_2021}.

% Experimental methods, however, have achieved high “re-identification” accuracy for datasets with numerous demographic attributes \cite{rocher_estimating_2019}. Marginalized individuals are especially vulnerable to these types of re-identification strategies, as there tend to be fewer data subjects in datasets that share their demographic attributes \cite{chang_privacy_2021}. Attempting to address this problem, researchers have proposed various differential privacy techniques for ensuring both a technical definition of fairness and non-identifiability, but these approaches are experimental and can inhibit other types of demographic analysis \cite{cummings_compatibility_2019,kuppam_fair_2020}.

Finally, another salient privacy risk is the possible loss of autonomy over one’s identity and interactions when demographic data is collected or used. Machine learning and AI systems are often built with the intention of making generalizations across groups in order to categorize individuals, meaning that it is not even necessary for an individual to share their demographic data in order for the system to decide to treat them as a “Black woman” or “Asian man.” Simply by matching patterns of behavior, algorithmic systems can categorize individuals, even if the categories are not explicitly labeled “Black woman” or “Asian man” \cite{mittelstadt_individual_2017,mavriki_automated_2019}. \citet{barocas_privacy_2019} refer to these types of associations between individuals as privacy dependencies, as an individual’s privacy quite literally depends on the privacy of the people like them. In other cases, even when users provide sensitive data about themselves, platforms may not take that data into account when making decisions for that user, subverting their agency around self-presentation \cite{bivens_gender_2017}.

For these types of privacy risks, we might expect privacy regulation such as GDPR or California’s California Consumer Privacy Act (CCPA) to prevent the worst abuses. Privacy regulation to date, however, has largely focused on the individual’s “right to privacy” and agency over their own personal data \cite{mittelstadt_individual_2017}, ignoring the relationality between data subjects and the interconnectedness of their privacy. Even when it comes to an individuals’ agency over data about them specifically, the relationship between individuals and the tech firms collecting their data is frequently one of “convention consent” \cite{taylor_public_2021}. In other words, users are resigned to share their data even when they do not agree with how it is being used because it is the cost of accessing platforms and services and they do not see any reasonable alternative \cite{draper_corporate_2019}. While there is technically always the option of not using platforms or services that require personal data, many have come to serve as essential infrastructure, calling into question how much someone can afford to hold onto their privacy by withholding their consent.

\subsection{Miscategorization and Identity Misrepresentation} \label{sec:indiv_misrep}

How demographic data is coded and represented in datasets — specifically, what categories are being used to define individual characteristics — can have a significant impact on the representation of marginalized individuals. When ADMS fail to accurately determine an individual's identity, such miscategorization and identity misrepresentation may not only lead to social and political discrimination, but also psychological and emotional harms via feelings of invalidation and rejection \cite{fosch-villaronga_little_2021}.

% In an effort to mitigate bias, some organizations seek to make their datasets more “representative” by including more data on different demographic categories such as race and gender. However, this is often done without a deeper engagement with the categories themselves or the collection methods used \cite{}. 

% As mentioned in \ref{sec:defdemo}, the demographic data required to enable current algorithmic fairness methodologies often does not account for the socially constructed nature of demographic categories, instead treating them as fixed, indisputable, apolitical attributes \cite{hanna_towards_2020, keyes_truth_2021, scheuerman_how_2020}. Race and gender are two demographic categories that are especially contextual, and debates and legislation around gender and race classification are constantly evolving \cite{roth_multiple_2016,kertzer_censuses_2001,lopez_whats_2021}. As such, it's important to recognize and account for the temporality of categorization during system design, as identity and categorization schemas can change over time.   

On one hand, individual miscategorization can occur when an individual is misclassified despite there being a representative category that they could have been classified under. To better understand the implications of miscategorization, it’s important to understand the different dimensions of identity and how these can lead to misrepresentation. With respect to racial identity, \citet{roth_multiple_2016} distinguishes between multiple dimensions of the concept of race, highlighting how an individual’s racial identity can be represented differently depending on the observer or method of data collection. Dimensions of racial identity include self-identity (the race an individual self-identifies as), self-classification (the racial category an individual identifies with on an official form), observed race (the race others believe you to be), appearance-based (observed race based on readily observable characteristics), interaction-based (observed race based on characteristics revealed through interaction such as language, accent, surname), reflected race (the race you believe others assume you to be), and phenotype (racial appearance) \cite{hanna_towards_2020}. When an individual is categorized under just one of these dimensions, unless use of the data is limited to a single bespoke purpose, it is highly likely that the individual will be misrepresented during disaggregated analyses in some way. For example, when one's self-classified race is collected and it differs from their most frequent perceived race, the analysis is likely to miss forms of discrimination stemming from perceived race \cite{lopez_whats_2021}. 

% For example, online social platforms that deploy recommendation algorithms may categorize individuals due to their perceived race (when self-reported race is not available), however, if an individual's perceived race is not aligned with their self-identity, the content they are recommended may not be as relevant to them or serve their experience on the platform. 

On the other hand, identity misrepresentation can occur when the categories used do not adequately represent an individual as they self-identify. As \citet{keyes_truth_2021} argues, ADMS designers and the data used carry particular expectations of what gender, class, or race mean in society. When categorization and classification of an individual is conducted by observation, either by person or machine, there is the risk that an individual's observed identity does not align with their self-identification and can lead to individual misrepresentation. Moreover, treating the notion of identity as a quality that can be “inferred” externally produces new forms of control over an individual’s agency to define themselves \cite{keyes_misgendering_2018, keyes_counting_2019, keyes_truth_2021}. For example, ADMS that involve predicting an individual's sexual identity perpetuate certain beliefs and ideas about queerness by associating specific characteristics, appearance, biology, or behavior as essential features of sexual identity \cite{tomasev_fairness_2021}. This can cause psychological harm to individuals who may not "fit the mold" of the category they self-identify with. As individuals come to understand the differences that form the basis for categorization, they can start to interpret their own actions through the lens of the category they are assigned to, in turn influencing their future decisions, a process that philosopher Ian Hacking dubbed the "looping effect" \cite{hacking_looping_1995}. For example, when individuals are made more acutely aware of what factors lead to them being perceived as “a woman” or as “queer,” they are incentivized to change their behavior either to increase the likelihood of their preferred classification or to simply live in a way they may now see as more aligned with their identity \cite{dembroff_real_2018}. Though this type of risk is not likely to be the most salient when collecting demographic data only to assess unequal outcomes or treatment, it is important to be mindful of when asking users for their demographics on platforms with content recommendations that are increasingly tailored to users based on the other information they provide, such as YouTube and TikTok.  

Additionally, restricting identity to fixed and measurable forms inherently misrepresents fluid and often unobserved characteristics such as sexuality and gender identity \cite{ruberg_data_2020, tomasev_fairness_2021}. Facial recognition technologies are a prominent case where the harm of identity misrepresentation occurs, since categorization is often based solely on observable characteristics. Many datasets used to train facial recognition systems are often built upon a binary, physiological perspective of female and male, and consequently misrepresent individuals who do not self-identify with those categories \cite{scheuerman_how_2020}. Continuing to build databases that assume identity is fixed and that only include observable traits risks reinforcing harmful practices of marginalization. Additionally, doing so can further entrench pseudoscientific practices which assume invisible aspects of one’s identity from visible characteristics such as physiognomy \cite{scheuerman_how_2020,stark_physiognomic_2021}. 

\subsection{Data Misuse and Use Beyond Informed Consent}
Once collected, demographic data can be susceptible to misuse. Misuse refers to the use of data for a purpose other than that for which it was collected or consent was obtained. In the context of ADMS, this could involve collecting and using data to train models that may be deployed in unexpected contexts or re-purposed for other goals. In practice, it is difficult for organizations to specify clear data uses at the point of collection. Sensitive demographic data, in this case, can go on to inform systems beyond the initial scope defined during collection. For example, in 2019 the U.S. government developed the Prisoner Assessment Tool Targeting Estimated Risk and Needs (PATTERN) \cite{us_department_of_justice_first_2019}. PATTERN was trained on demographic data and criminal history data for the purpose of assessing recidivism risk and providing guidance on recidivism reduction programming and productive activities for incarcerated people \cite{us_department_of_justice_first_2019}. Then, in March 2020 the Bureau of Prisons was directed to begin using PATTERN to determine which individuals to transfer from federal prison to home confinement in the wake of the COVID-19 pandemic \cite{partnership_on_ai_algorithmic_2020}. However, the data used to inform PATTERN was not intended to inform inmate transfers, let alone during a global pandemic which introduced a number of unprecedented social and economic variables \cite{partnership_on_ai_algorithmic_2020}.

Data misuse could also refer to instances where data is shared with third parties or packaged and sold to other organizations. A notable example of data misuse in this respect can be seen in Clearview AI’s facial recognition dataset, which the company claims contains over three billion images scraped from social media platforms such as Facebook, Instagram, LinkedIn, and Twitter, along with personal information listed on people’s social media profiles \cite{hill_secretive_2020}.

Corporations collecting and using individuals' demographic data to train and deploy ADMS are facing some increased pressure (from both the public and regulatory bodies) for transparency on how such data is collected and used. For example, Article 13 of the GDPR \cite{european_parliament_and_council_of_european_union_regulation_2016} requires companies collecting personal data from a data subject to provide the data subject with information such as the purpose of the data processing, where the data is being processed and by which entity, recipients of the data, the period for which the data will be stored, the existence of algorithmic decision-making and the logic involved, and the right to withdraw data \cite{european_parliament_and_council_of_european_union_regulation_2016}. Companies have begun to incorporate this informational requirement into their data collection practices, often in the form of click wraps, digital banners that appear on users’ screens and require them to “accept all” or “decline” a company’s digital policies. Yet, providing individuals with transparency and information about how data will be used is generally not sufficient to ensure adequate privacy and reputational protections \cite{obar_sunlight_2020}. Overloading people with descriptions of how their data is used and shared and by what mechanisms is not a way to meaningfully acquire data subjects’ consent, especially in cases where they are sharing sensitive, personal information \cite{obar_sunlight_2020,oeldorf-hirsch_overwhelming_2019}. Rather, the goals of data use and the network of actors expected to have access to the data are what need to be clearly outlined and agreed upon by the data subject. Additionally, while it may be difficult for organizations to specify clear data uses at the point of collection, companies may consider providing updates as the use cases for that data becomes clearer. In following with this more rigorous notion of consent, we would expect check-ins on how the data was used to assess or mitigate discrimination and on whether the data subjects would still like for their sensitive data to be used towards these ends. Collecting demographic data consensually requires clear, specific, and limited use as well as strong security and protection following collection.

\section{Community Risks of Demographic Data Collection and Use} \label{sec:commrisks}
Moving beyond individual risks, this section considers a range of potential harms to communities. As ADMS seek to generalize across groups based on data collected from a subset of the population, data collection can lead to a number of unintended risks including undue surveillance, group misrepresentation, and the ceding of agency over defining what constitutes fair and just treatment, which we detail below. 

\subsection{Expanding Surveillance Infrastructure in the Pursuit of Fairness} \label{sec:surveillance}
Data collection is now employed on a regular basis to define and monitor types of groups, such as customers, communities, or populations. This type of surveillance does not target individuals directly, but looks at how people can be grouped together or what it means to be a member of a specific group. This form of group profiling raises a number of questions around harms related to collective privacy and discrimination \cite{mittelstadt_individual_2017,mavriki_automated_2019}.

As discussed throughout this paper, there is often a trade-off between privacy and fairness when it comes to assessing discrimination and inequality. Calls to collect demographic data in order to enable algorithmic fairness techniques run the risk of intersecting with many corporations' and governments' attempts to understand the impacts their products and services are having on groups at large, potentially justifying the expansion of data-driven surveillance infrastructures. Scholars of surveillance and privacy have shown time and time again that the most disenfranchised and “at-risk” communities are routinely made “hypervisible” by being subjected to invasive, cumbersome, and experimental data collection methods, often under the rationale of improving services and resource allocation \cite{benjamin_race_2019,browne_dark_2015,eubanks_automating_2017}. Within this context it is not unreasonable for members of disenfranchised groups to distrust new data collection efforts and to withhold information about themselves when sharing it is optional. 

Further still, increased visibility and awareness of being under surveillance is likely to have a chilling effect on community groups and society at large. \citet{citron_privacy_2021} highlights that data-based surveillance can reduce the range of viewpoints and amount of information shared among communities. One example of this is the dramatic decrease of Grindr users sharing their HIV status on the app when it was learned that Grindr had shared this data with analytics firms \cite{citron_privacy_2021}. In this way, attempts to gain insight into specific groups through demographic data collection may result in widespread self-censoring. Though most practitioners are well-meaning in their efforts to improve representation and system performance for groups, it is important to consider what the cost and risks of being better represented in datasets towards underrepresented groups \cite{hoffmann_terms_2020}.

In cases where there seems to be a trade off between institutional visibility or anti-discrimination and surveillance, we recommend centering the agency of the groups that planned interventions are supposed to support. Scholarship from the emerging fields of Indigenous Data Sovereignty and Data Justice can provide a starting point for what this might look like — instead of collecting demographic data to “objectively” or “authoritatively” diagnose a problem in the system or even in society more broadly, data collection efforts can be grounded in community needs and understandings first and foremost \cite{rainie_indigenous_2019, ricaurte_data_2019, walter_delivering_2020}. It is also important to note that disaggregated data is not the only way that groups facing discrimination or other forms of inequality can become more visible. Small-scale data collection and qualitative methodologies can also be used to identify treatment and outcome disparities \cite{rosen_racial_2021,massey_riding_2016,welles_minorities_2014}. 

\subsection{Group Misrepresentation and Reinforcing Oppressive or Overly Prescriptive Categories} \label{sec:group_misrep}
Another source of risk arises from the demographic categories themselves and what they are taken to represent. Scholars from a wide range of disciplines have considered the question of what constitutes representative or useful categorization schemas for race, gender, sexuality, and other demographics of institutional interest and where there are potential sources for harm \cite{bowker_sorting_1999,dembroff_real_2018,hacking_looping_1995,hanna_towards_2020,hu_what_2020,keyes_counting_2019,zuberi_white_2008,guyan2022queer}. Though there are certainly nuances to defining and measuring each of these demographics, we can find some general trends across this scholarship around the risks of uncritically relying on these categories to describe the world, or, in our case, to ascertain treatment differences across groups. At a high level, these risks center around essentializing or naturalizing schemas of categorization, categorizing without flexibility over space and time, and misrepresenting reality by treating demographic categories as isolated variables instead of “structural, institutional, and relational phenomenon” \cite[1]{hanna_towards_2020}.

The first of these risks, and certainly the one most frequently encountered and vocalized by practitioners \cite{andrus_what_2021}, is when entire groups are forced into boxes that do not align with or represent their identity and lived experience. Often, this occurs because the range of demographic categories is too narrow, such as leaving out options for “non-binary” or “gender-fluid” in the case of gender \cite{bivens_gender_2017}. It can also commonly occur in cases where demographic data is collected through inference or ascription. In these cases, systems often embed very narrow standards for what it means to be part of a group, defining elements of identity in a way that does not align with the experience of entire segments of the population \cite{fosch-villaronga_little_2021}. This type of risk is especially well-documented with regards to various types of automated gender recognition failing to correctly categorize transgender and non-binary individuals. Both critics and users deem these failures inevitable because these systems treat gender as purely physiological or visual, which is different from how members of these communities actually experience gender \cite{hamidi_gender_2018,keyes_misgendering_2018,keyes_truth_2021, fosch-villaronga_little_2021}. In each of these ways, demographic data collection efforts can reinforce oppressive norms and the delegitimization of disenfranchised groups, potentially excluding entire communities from services and institutional recognition as a form of what critical trans scholar Dean Spade calls “administrative violence” \cite{spade_normal_2015}. 

Furthermore, data collected with too limited of categories risks misrepresenting and obscuring subgroups subject to distinct forms of discrimination and inequality, especially in cases where demographic data is collected via inference. The most common inference techniques used by public and private institutions generally rely on the very limited set of demographic categories included in the census, such as Bayesian Improved Surname Geocoding (BISG), which uses an individual’s name and zip code to predict their race \cite{elliott_using_2009}. As one example, there have been many efforts to distinguish between Asian American and Pacific Islander (AAPI) populations in health \cite{shimkhada_capturing_2021} and education \cite{poon_count_2017} due to fears that disenfranchised subgroups are made further invisible by being categorized under the broad umbrella of AAPI. Models like BISG, however, use U.S. census data and thus cannot go beyond the six census categories for race and ethnicity (White, Black, AAPI, American Indian/Alaskan Native, and Multiracial).

Another way that categorization schema can be misaligned with various groups’ experiences and lived realities is when the demographic categories themselves are too narrowly defined to capture all the dimensions of possible inequality. For example, each of the dimensions of race discussed in Section \ref{sec:indiv_misrep} carries with it different potential adverse treatments and effects. If the only type of demographic data an institution collects is through self-identification, for instance, it can draw a very different picture of discrimination than data collected through ascription \cite{hanna_towards_2020,saperstein_capturing_2012}. As such, when it comes to assessing discrimination or some other form of inequality, it is critical that practitioners have a prior understanding of how differential treatment or outcomes are likely to occur such that the right dimension of identity is captured to accurately assess likely inequities.

As previously mentioned, it is also important to consider the temporality of categorization — categorization schema and identities can change over time, and how much this is taken into account during system design will likely have a disproportionate impact on groups with more fluidity in their identities. Looking first to gender and sexuality, critical data scholars have argued that queer and trans identities are inherently fluid, contextual, and reliant upon individual autonomy \cite{keyes_counting_2019,ruberg_data_2020}. There are no tests or immutable standards for what it means to be queer, non-binary, or any number of other forms of identity, and it is likely that one’s presentation will change over time given new experiences and contexts. In other words, queer identities can be seen as perpetually in a state of \textit{becoming}, such that, rigid, persistent categorizations into states of being can actually be antithetical to these identities. Pushing towards actionable interventions, \citet{tomasev_fairness_2021} suggest moving past attempts to more accurately label queer individuals and groups as a way of achieving fairness and looking instead to qualitatively engage with queer experiences with platforms and services to see how cisheteronormativity crops up in system design.

Somewhat similarly, in studies of race and racism it has been argued that race should be seen as a “a dynamic and interactive process, rather than a fixed thing that someone has” \cite{pauker_review_2018}. Especially for multiracial individuals, there is immense malleability in how they are perceived by others, how they perceive themselves, and what they choose to accentuate in their presentation and interactions to influence various forms of racial classification \cite{pauker_review_2018}. Similar to the case with queer identities, attempts to develop and enforce fairness constraints around more static, decontextualized notions of race will miss the ways in which forcing groups into static boxes is itself a form of unfairness. As such, when it is not possible to work with these fluid identity groups directly to understand how systems fail to accommodate them, data subjects should at the very least be given opportunities to update or clarify their demographics in cases where data is collected over an extended period of time and it is used in variable contexts \cite{ruberg_data_2020}.

Even in cases where groups feel adequately represented by a categorization schema, however, the categories can become harmful depending on how they are used. When demographic categories start to form the basis for differences in servicing, such as in advertising and content recommendation, there is a risk of reinforcing and naturalizing the distinctions between groups. Especially in cases where demographic variables are uncritically adopted as an axis for differential analysis, varying outcomes across groups can be incorrectly attributed to these variables, as has occurred many times in medical research \cite{braun_racial_2007}, which in turn reinforces the notion that the differences between groups are natural and not a result of other social factors \cite{hanna_towards_2020}. With regards to race, a categorization schema that is conclusively not genetic or otherwise biological \cite{morning_does_2014}, this has been described as the risk of studying race instead of racism. By looking for differences between what groups do instead of how groups are treated, it encourages attributing responsibility to oppressed groups for their own oppression. For example, in the creation of recidivism risk scores tools for the criminal justice system, there has been extensive focus on what factors increase the accuracy of criminality prediction \cite{barabas_beyond_2019}. However, given how criminality is usually defined — namely, that an individual has been arrested and charged for a crime — the factors that end up predicting criminality most accurately are often just the factors that increase one’s likelihood to be targeted by discriminatory policing \cite{barabas_beyond_2019}.

\subsection{Private Control Over Scoping Bias and Discrimination} \label{sec:priv_fair}
As a final risk to consider, the assessment of inequality and discrimination is a not rigidly defined or widely agreed upon process. Rather, institutions that collect demographic data have a wide range of techniques and approaches they can possibly employ when it comes to both collecting data and interpreting that data. As such, if an institution is asking already marginalized groups to share information for the purposes of assessing unfairness, it is imperative for that institution to operationalize fairness in a way that is aligned with these groups' interests.

In determining what standards of fairness an institution is likely to use, it can be instructive to consider the institution’s motivations for conducting measurements of fairness in the first place. Though there are many reasons an institution might try to assess and mitigate discrimination and inequalities in their machine learning and algorithmic decision-making systems, much of this work is motivated at least in part by concerns around liability \cite{andrus_what_2021,holstein_improving_2019,rakova_where_2021}. Generally speaking, however, legal notions of discrimination and fairness remain somewhat limited, often esteeming “neutral” decision-making that attempts to treat everyone the same way as the path towards equality \cite{wachter_bias_2021,xenidis_tuning_2021,xiang2020reconciling}. As such, most deployed methods in the algorithmic fairness space are geared towards “de-biasing” decision-making to make it more neutral, rather than trying to directly achieve equality, equity, or another form of social justice \cite{balayn_beyond_2021}. Given disparate starting points for disenfranchised groups, however, this view that neutrality can lead to a more equal world is both risky and unrealistic, as attempts to be neutral or objective often have the effect of reinforcing the status quo \cite{fazelpour_algorithmic_2020,green_algorithmic_2020}. Despite this, commitments to neutrality remain the norm for many governmental and corporate policies. 

Another element of most technical approaches to fairness measurement is that they are strictly formalized. Formalizability refers to the degree to which it is possible to represent a definition of fairness through mathematical or statistical terms — for instance, defining fairness as correctly positively categorizing individuals from different groups at the same rate (i.e. equality of true positive rates \cite{mitchell_algorithmic_2020}) is distinctly formalizable. Formalizability is an important attribute of fairness when it has to also coincide with the system design values of efficiency and scalability, because formalization enables a system designer to treat many different problems (e.g. racism, sexism, ableism) similarly. That being said, it also relies on treating much of the world as static. As \citet{green_algorithmic_2020} have argued, by treating the point of decision-making as the only possible site of intervention (i.e. adjusting predictions to adhere to some notion of fairness), these attempts at formalization hold fixed many of the engines of discrimination, such as the ways in which different groups interact with institutions and why differences might exist between groups in the first place.

Just as defining fairness, discrimination, or bias is impacted by an institution’s goals and values, the collecting, processing, and interpreting of data is never truly objective. In other words, data is never "raw" because it is shaped by the conditions in which it was collected, the methods that were used, and the goals of measuring the world in the first place \cite{gitelman_raw_2013}. This brings up a salient source of risk in the collection of demographic data: the types of discrimination and inequality that can be assessed using demographic data are largely determined by what other types of data are being collected. For instance, it might be possible to detect that a risk score recidivism tool has unequal outcomes for members of different groups, but without accurate data about interactions between suspects or defendants and police and judges, it may not be possible to accurately assess why these inequalities show up in the data and thus how to best address them \cite{barabas_studying_2020}. Given that data collection efforts must be consciously designed, data always reflects some viewpoint on what is important to understand about the world. When those collecting data have blindspots about what impacts decision-making and individuals’ life experiences, various forms of discrimination and inequality run the risk of being misread as inherent qualities of groups or cultural differences between them \cite{crooks_numbers_2021}. This is why historian Khalil Gibran has argued that the seemingly objective focus on data and statistical reasoning has replaced more explicitly racist understandings of racial difference \cite{muhammad_condemnation_2019}, a shift made possible by disaggregated data.

Taking these subjectivities of fairness measurement into account, there is a significant risk that the collection of demographic data enables private entities to selectively tweak their systems and present them as fair without meaningfully improving the experience of marginalized groups. So long as the data used to assess fairness is collected and housed by private actors, these actors are given substantial agency in scoping what constitutes fair decision-making going forward. One striking example of this already occurring is the creation and normalization of “actuarial fairness,” or the notion that “each person should pay for his own risk,” in the insurance industry \cite{ochigame_beyond_2018}. Using statistical arguments about the uneven distribution of risk across different demographic categories, industry professionals were able to make the case for what previously might have been considered outright discrimination — charging someone more for insurance because their immutable demographic categorizations increase their statistical risk \cite{ochigame_beyond_2018}.

Finally, though this work is motivated by the documented unfairness of ADMS, it is critical to recognize that bias and discrimination are not the only possible harms stemming directly from ADMS. As recent papers and reports have forcefully argued, focusing on debiasing datasets and algorithms can draw attention away from other, possibly more salient harms \cite{balayn_beyond_2021}. For many ADMS that are clearly susceptible to bias, the greater source of harm could arguably be the deployment of the system in the first place \cite{milner_abolish_2019,barabas_build_2020}. Attempting to collect demographic data in these cases will likely do more harm than good, as demographic data will draw attention away from harms inherent to the system and towards seemingly resolvable issues around bias.

\section{Paths Forward} \label{sec:paths}

\subsection{Anonymity, Cryptographic Privacy, and Third Party Data Management} \label{sec:crypto}
In response to many of the measurement concerns in Section \ref{sec:measurement}, there have been a range of proposals on how to feasibly collect, manage, and employ demographic data without corporations ever directly learning users' sensitive attributes. Using technical methods that enable trusted third parties to be the data collectors and holders, researchers and practitioners have found ways to maintain individual non-identifiability throughout the data's use \cite{veale_fairer_2017,kilbertus_blind_2018}. 

Most of these approaches prioritize anonymizing datasets that include demographics by enforcing k-anonymity, p-sensitivity, and/or differential privacy. Ensuring k-anonymity involves narrowing down the fields in a dataset and lumping variable ranges together in order to ensure that no individual has a unique set of values in the dataset that they might be re-identified with \cite{sweeney_k-anonymity_2002}. P-sensitivity involves perturbing sensitive attribute responses such that even if you know all the other variables for an individual member of the dataset, you would not be able to concretely determine the individual's sensitive attribute \cite{basu_measuring_2020}. Differential privacy, unlike the previous two strategies, focuses instead on the model or analysis resulting from the use of sensitive data and ensures that the model or analysis would be unchanged by the removal of an individual, such that it is impossible to tell post-processing whether a specific individual was included or not in the dataset \cite{jagielski_differentially_2019}. It is important to note that all of these anonymization approaches, however, add their own sources of bias that can generate misleading conclusions \cite{kuppam_fair_2020,hajian_study_2012}. 

Another range of strategies employ secure, multiparty computation (SMPC) as a means of protecting sensitive attribute data from the primary institution and outside attackers. This approach ensures individual sensitive attribute data remains encrypted at each stage of use, while still being able to carry out basic computations with the encrypted data to generate aggregate level takeaways \cite{alao_how_2021,kilbertus_blind_2018}.

What these methods often neglect, however is the range of individual harms that extend from the relational aspects of data and the community level risks of demographic data collection. Secure computation techniques are generally concerned with an "identifiability" notion of privacy which centers the question of if data is attributable to an individual. What they miss is a more "control" centric understanding of privacy, which speaks to the ability of individuals to influence what data about them exists and how it gets used \cite{viljoen_democratic_2020}. Of the risks discussed in sections \ref{sec:indivrisk} and \ref{sec:commrisks}, anonymization and cryptographic privacy can protect against the direct discrimination of having one's sensitive attributes revealed and some types of data misuse and surveillance, but they do not address concerns of (mis)representation and loss of agency. As such, we turn to recent proposals around forms of participatory data governance as a possible mechanism for mitigating these risks.

\subsection{Participatory Governance of Sensitive Data} \label{sec:part_gov}
Data governance is an increasingly popular topic of discussion in light of the ever-growing swaths of data held by governments and corporations used without citizen or consumer accountability. Borrowing from \citet{micheli_emerging_2020}, data governance can be described as "\textit{the power relations between all the actors affected by, or having an effect on, the way data is accessed, controlled, shared, and used, the various socio-technical arrangements set in place to generate value from data, and how such value is redistributed between actors}." We argue that emerging data governance models which move away from individualistic data rights towards collective forms of data governance \cite{viljoen_democratic_2020} can help mitigate the risks to individuals and communities described above and enable the responsible collection and use of demographic data. 

While a range of alternative data governance models have been proposed \cite{van_geuns_what_2020}, data cooperatives and data trusts hold notable promise for overcoming both the individual and community risks outlined in this paper. Data cooperatives are characterized by a "de-centralized data governance approach in which data subjects voluntarily pool their data together to create a common pool for mutual benefits" \cite{ho_governance_2019}. Data trusts, on the other hand, generally make use of a more centralized structure that relies on an independent data "trustee" to steward over the voluntary pooling and external sharing or contracting of data \cite{aapti_institute_enabling_2021}. Notably, these forms of data governance addresses the power imbalances characteristic of current data regimes by prioritizing the distribution of access and rights to data across its members. 

Looking back to the range of risks described in sections \ref{sec:indivrisk} and \ref{sec:commrisks}, these types of participatory data governance strategies can help address each in unique ways. In terms of the individual risk of privacy, participatory governance strategies inherently offer greater control to individuals over what is known about them by whom. And though nothing in the definition of data trusts or data cooperatives requires data to be nonidentifiable, both governance strategies could be used to establish the type of trusted third party relationships described in the previous section to enable encrypted or anonymized data use. In this way, we do not believe that participatory governance strategies are an alternative to cryptographic privacy and third-party data arrangements, but that they can actually be complementary strategies in mitigating privacy risks. These types of data governance models also have the potential to mitigate the risks of individual miscategorization and identity misrepresentation due to involvement of the data subjects in decisions regarding what types of data are collected and how these are categorized, and ability to raise concerns through deliberations or with data stewards. Therefore, it's possible to imagine that a participatory data governance model would include categories that more accurately represent the individuals it seeks to benefit. Similarly, the risks of data misuse or use beyond informed consent are also greatly mitigated given that it is one of the key focuses of participatory data governance to have clear, concrete boundaries around what data is collected, why it's collected, and how it's used.
    
Turning to the community level risks of demographic data collection, the situation becomes more nuanced. The threat of expanding surveillance infrastructure, while reduced, is not altogether mitigated by the types of governance structures we have described or by voluntary participation in data collection efforts more broadly. Depending on how individuals are incentivized to contribute their data to a data trust or data cooperative, there is some threat of already exploited and disenfranchised populations disproportionately offering up their data for financial or material benefits, reinforcing existing disparities in privacy \cite{marwick_privacy_2018,vasquez-tokos_racialization_2021}. Furthermore, surveillance operates on a community scale, not just at the level of the individual. Given the relational nature of data, data freely shared with a data trust or cooperative that is then used by private entities to build ADMS or other types of machine learning models has implications for those that have not shared their data as well. As seen through Immigration and Customs Enforcement's (ICE) purchase and use of mobile phone location data to model the movement patterns of "undocumented immigrants" \cite{tau_federal_2020}, access to the sensitive attributes of some individuals (e.g. documentation status), can enable the surveillance of a much larger group. As such, it is incumbent upon the data trustees and data cooperatives to stay aware of how the data they wield carries risks for individuals outside their organization and to mitigate these risks where they can. 

Participatory governance models can also provide a point of intervention to addressing the risk of group misrepresentation and the reinforcement of oppressive categories by enabling previously disenfranchised groups to directly define their group identity and exercise control over the applications of their community's data. One example where the utility of this has already been seen is in the field of Indigenous Data Sovereignty, which centers Indigenous peoples' prerogative to govern the collection, access, and use of their data. In cases where Indigenous tribes have lead their own data collection efforts, they have pushed back against external categorization schemas of determining tribal citizenship, such as externally imposed standards of "blood quantum," in order to more accurately define tribal membership and tell the story of their tribes \cite{rainie_data_2017}.  Similarly, participatory governance models give data subjects the ability to exercise influence over how fairness is defined, what types of fairness assessments should be conducted, and what data should be used to conduct those assessments in a given context. This reduces the control private actors have over the operationalization of fairness and increases the alignment of fairness objectives with data subjects interests and their desired interactions with specific ADMS. However, realizing this potential will likely require some expertise around algorithmic fairness or discrimination from within the data governance structure.

While participatory data governance strategies present an opportunity to mitigate many of the risks and harms discussed in this paper, the feasibility concern from Section \ref{sec:measurement} still looms large. If the builders of ADMS are not externally compelled to approach demographic data collection and use responsibly, or even to engage with anti-discrimination in the first place, it is unlikely that data cooperatives or trusts form in order to responsibly manage and provide this data. As such, getting there will require the efforts of practitioners on the inside making the case for working with external data governance structures and of academics, policy-makers, and activists on the outside pushing for the enforcement of anti-discrimination standards and data privacy protections that would encourage corporations to only access sensitive demographic data through these limited means. As companies like Meta and AirBnB start to explore third party data holder arrangements for fairness assessments, however, this future starts to look more feasible \cite{alao_how_2021,basu_measuring_2020}.

%%diffuclty of ensuring data is used for the purpose it was initially collected for

\begin{acks}
We are grateful to the diverse set of individuals who engaged with us over the last year through one-on-one calls as well as the PAI-hosted FAccT CRAFT workshop and RightsCon session. We would like to thank our colleagues Christine Custis and Tina Park who provided advice and feedback on drafts of this paper. While this document reflects the input of individuals representing many PAI Partner organizations, it should not be read as representing the views of any particular organization or individual or any specific PAI Partner.

\end{acks}

\section*{Funding Disclosure}
Funding for this study was provided by Partnership on AI. Partnership on AI is funded by a combination of philanthropic institutions and corporate charitable contributions. Primary corporate funding is always considered general operating support and legally classified as non-earmarked charitable contributions (not donations in exchange for goods or services, or quid pro quo contributions) to avoid the possibility of conflict in corporate funders having undue influence on Partnership on AI’s agenda or on particular programs. More detail on Partnership on AI’s \hyperlink{https://partnershiponai.org/transparency-governance/}{funding and governance} is available online.

\bibliographystyle{ACM-Reference-Format}
\bibliography{demodata}

% \appendix



\end{document}